\documentclass[%
 reprint,
superscriptaddress,
 amsmath,amssymb,
 aps,prl
]{revtex4-1}

\usepackage{graphicx}
\usepackage{dcolumn}
\usepackage[utf8]{inputenc}
\usepackage{bm}
\usepackage[colorlinks=true,linkcolor=blue,citecolor=blue]{hyperref}
\usepackage{float}

\renewcommand{\v}{\bm{v}}

\usepackage{amsmath,bm}
\renewcommand{\tilde}[1]{\widetilde{#1}}
\DeclareMathAlphabet\mathbfcal{OMS}{cmsy}{b}{n}
\usepackage{lipsum}

\usepackage[normalem]{ulem}

\usepackage{xcolor}

\newcommand{\figref}[1]{Fig. \ref{#1}}

\begin{document}



\title{On the Cascade-Dissipation Balance in Astrophysical Plasmas}
\author{D. Manzini}
\email{davide.manzini@lpp.polytechnique.fr}
\affiliation{Laboratoire de Physique des Plasmas (LPP), CNRS, École Polytechnique, Sorbonne Université, Université Paris-Saclay, Observatoire de Paris, 91120 Palaiseau, France}%
\affiliation{Dipartimento di Fisica E.Fermi, University of Pisa, Italy}%

\author{F. Sahraoui}
\affiliation{Laboratoire de Physique des Plasmas (LPP), CNRS, École Polytechnique, Sorbonne Université, Université Paris-Saclay, Observatoire de Paris, 91120 Palaiseau, France}%

\author{F. Califano}
\affiliation{Dipartimento di Fisica E.Fermi, University of Pisa, Italy}%

\date{\today}

\begin{abstract}
The differential heating of electrons and ions by turbulence in weakly collisional magnetized plasmas and the scales at which such energy dissipation is most effective are still debated. Using a large data sample measured in the Earth's magnetosheath by the Magnetospheric Multiscale mission and the coarse-grained energy equations derived from the Vlasov-Maxwell system we find evidence of a balance over two decades in scales between the energy cascade and dissipation rates. The decline of cascade rate at kinetic scales (in contrasts with a constant one in the inertial range), is balanced by an increasing ion and electron heating rates, estimated via the pressure-strain. Ion scales are found to contribute most effectively to ion heating, while electron heating originates equally from ion and electron scales. These results can potentially impact current understanding of particle heating in turbulent magnetized plasmas as well as their theoretical and numerical modeling.

\end{abstract} 

\maketitle

\noindent One of the central problems in turbulent media is to understand how energy is transferred across scales and  how it is eventually dissipated. For weakly collisional plasma such as the solar wind or planetary magnetosheahts the pioneering work of \citet{politano1998} enable us, in the framework of incompressible Magnetohydrodynamics (MHD) and under the assumptions of fully developed turbulence, to express the cascade rate as a function of third order structure functions of the velocity and magnetic field \citep{sahraoui2020}. These results have been extended to account for compressibility \citep{banerjee2013,andres2018}, the contribution of the Hall current at subion scales \citep{galtier2008,hellinger2018,ferrand19}, different fluid closure equations \citep{simon2021} and temperature anisotropy \citep{simon2022} and have been used extensively to measure the cascade rate in spacecraft data \cite{sorriso-valvo2007, MacBride_2008,Stawarz_2009,stawarz10,Coburn_2012,hadid17,hadid18,Bandyopadhyay_2018,andres2019,Bandyopadhyay_2020,andres_2021,brodiano23,pecora23}.
\noindent In recent years the coarse-graining (CG) method, initially developed for hydrodynamics \citep{germano_turbulence_1992,eyink_locality_2005}, gained popularity in the plasma physics community \citep{aluieMHD,CamporealePRL,Cerri_camporeale,Yang17,manzini2022}. This formulation provides an alternative way to measure the cascade rate and enables us to overcome some limitations imposed by the stringent hypotheses of fully developed turbulence (e.g., spatial homogeneity).
Indeed, the CG approach can be employed not only to measure the average cascade rate over large plasma portions, but also to address localized cross-scale energy transfer in reconnecting current sheets \citep{manzini23, adhikari2023scale}. In this work we scale-filter the Vlasov-Maxwell system of equations and measure the nonlinear energy cascade rate and the exchanges between its various forms (kinetic, electromagnetic and thermal) as a function of scale, with particular focus on the turbulent plasma heating given by the pressure-strain interaction \citep{Belmont_book,Fiitzpatrick_book}.\\
\noindent \textit{The Coarse Graining theory--} To study cross-scale energy transfer we apply the spatial CG approach to the moments of Vlasov equation, written for an electron-ion plasma ($\alpha=e,i$) , and the Maxwell ones. All variables are low-pass filtered at a scale $\ell$, \emph{e.g} $\overline{\bm{v}}_\ell=\int d\bm{r}G_\ell(\bm{r})\bm{v}(\bm{x}+\bm{r})$, where $G_\ell$ is a centered, normalized filtering kernel with variance of order $\ell^2$. To include the effects of compressibility we introduce a density weighted filtering (\emph{Favre filtering}) defined for a given field $f$ as: ${\overline{\rho}_\ell}\tilde{{f}}_\ell={\overline{\rho {f}}_\ell}$ \citep{aluie_compressible_2011,Aluie_compressible}. For conciseness of the notations the filtering scale $\ell$ is not written explicitly unless necessary.\\

\noindent At each scale $\ell$ we can write the equations for the large-scale bulk flow ($\widetilde{\mathcal{E}}^{\mathrm{f}}_\alpha=\overline{\rho_\alpha}|\tilde{\v_\alpha}|^2/2$), electromagnetic (EM, $\overline{\mathcal{E}}^{em}\!=\!({|\overline{\bm{E}}|^2+\!|\overline{\bm{B}}|^2})/{8\pi}$) and thermal energies ($\overline{\mathcal{E}}^{\mathrm{th}}_\alpha=\mathrm{Tr}(\overline{\bm{P}_\alpha})/2$):
\begin{eqnarray}
\frac{\partial}{\partial t} \left( \widetilde{\mathcal{E}}^{\mathrm{f}}_i+\widetilde{\mathcal{E}}^{\mathrm{f}}_e \right)&=&-\nabla\cdot{\mathbfcal{F}}^{\mathrm{f}}_\ell+\overline{\bm{j}}\cdot\overline{\bm{E}} \nonumber\\&+&\overline{\bm{P}_i}:\nabla\overline{\bm{v}_i}+\overline{\bm{P}_e}:\nabla\overline{\bm{v}_e} -\pi(x,\ell) \label{eq_k}\\
\frac{\partial}{\partial t} \overline{\mathcal{E}}^{em}&=&-\nabla\cdot\mathbfcal{F}^{\mathrm{em}}_\ell-\overline{\bm{j}}\cdot\overline{\bm{E}}\label{eq_em}\\
\frac{\partial}{\partial t} \overline{\mathcal{E}}^{\mathrm{th}}_\alpha&=&-\nabla\cdot\mathbfcal{F}^{\mathrm{th}}_\ell-\nabla\cdot\overline{\mathbf{h}}_\alpha\nonumber \\&-&\overline{\bm{P}_\alpha}:\nabla\overline{\bm{v}_\alpha} -\phi_\alpha(x,\ell)\label{eq_th}
\label{eq:energies}
\end{eqnarray}
where
\begin{equation}
\begin{split}
\pi(x,\ell)= &\sum_{\alpha=e,i} -\overline{\rho_\alpha}\left[\widetilde{\v_\alpha\v_\alpha}-\tilde{\v_\alpha}\tilde{\v}_\alpha\right]:\nabla\tilde{\v_\alpha}\\
&+(\nabla\overline{\bm{P}_\alpha})\cdot(\tilde{\v_\alpha}-\overline{\v_\alpha})\\
& - \overline{n_\alpha q_\alpha \v_\alpha}\cdot\left[ (\tilde{\bm{E}}-\overline{\bm{E}}) + \frac{1}{c}\left(\widetilde{\v_\alpha\times\bm{B}} - \tilde{\bm{v}_\alpha}\times\tilde{\bm{B}}\right) \right]
\end{split}
\end{equation}
is the cross-scale energy transfer (or \emph{turbulent cascade}) rate  across the scale $\ell$.
The quantities $\overline{\bm{j}}\cdot\overline{\bm{E}}$ and $\mathrm{PS}_\alpha\equiv\overline{\bm{P}_\alpha}:\nabla\overline{\bm{v}_\alpha}$ are the only sink terms of the large-scale EM and Thermal energy, respectively, and appear as a source in the large scale bulk flow energy. This implies that any process that changes the large scale thermal energy must go through the $\mathrm{PS}_\alpha$ channel. At the same time equation \eqref{eq_k} shows that (a fraction of) $\bm{j}.\bm{E}$ can lead to plasma heating (via the $\mathrm{PS}_\alpha$) without having to modify the large scale bulk flow energy. This can be seen even more clearly by summing equations (\ref{eq_k}) and (\ref{eq_th}), which indicates that $\bm{j}\cdot\bm{E}$ acts as a source for the total kinetic energy of the plasma particles (bulk flow {\it and} thermal energy) \citep{howes_klein_li_2017,afshari_ELD}.
The quantity $\phi_\alpha=\overline{\bm{P}_\alpha:\nabla\bm{v}_\alpha}-\overline{\bm{P}_\alpha}:\nabla\overline{\bm{v}_\alpha}$ stands for a nonlinear cascade of thermal energy. This term transfers thermal energy from large to small spatial scales and will not be discussed in this work since we are interested only in the transfer to the thermal energy. While $\overline{\bm{j}}\cdot\overline{\bm{E}}$ and $\overline{\bm{P}_\alpha}:\nabla\overline{\bm{v}_\alpha}$ are cumulative quantities, encompassing the energy exchanges from all scales larger than $\ell$, the cross-scale terms $\pi,\phi$ measure the transfer \emph{across} scale $\ell$.
The spatial fluxes in the form $\nabla\cdot\mathbfcal{F}$, including the divergence of the filtered heat flux $\nabla\cdot\overline{\mathbf{h}_\alpha}$, move the large scale energies in space and disappear after integration over a suitable domain.   \\
A similar set of equations was derived in \citep{Yang17}. In this work the cascade rate includes the term $\nabla \overline{\bm{P}}_\alpha\cdot(\tilde{\bm{v}_\alpha}-\overline{\bm{v}_\alpha})$, also present in compressible hydrodynamics \citep{aluie_compressible_2011}.\\
Summing equations (\ref{eq_k})-(\ref{eq_em}) and averaging over a portion of plasma yields :
\begin{eqnarray}
\frac{\partial}{\partial t}\left\langle\widetilde{\mathcal{E}}^{\mathrm{f}}_i+\widetilde{\mathcal{E}}^{\mathrm{f}}_e+\overline{\mathcal{E}}^{em}\right\rangle + \nabla\cdot\left\langle  {\mathbfcal{F}}^{\mathrm{f}}_\ell+{\mathbfcal{F}}^{\mathrm{em}}_\ell \right\rangle&=& \nonumber \\ \mathrm{PS}_i(\ell)+\mathrm{PS}_e(\ell)&-&\Pi(\ell)
\label{eq:balance}
\end{eqnarray}
where in the right hand side we find the average $\mathrm{PS}$ interaction, $\mathrm{PS}_\alpha(\ell)=\langle\overline{\bm{P}_\alpha}:\nabla\overline{\bm{v}_\alpha}\rangle$, filtered at scale $\ell$, and the net energy cascade $\Pi(\ell)=\langle\pi_\ell\rangle$.\\
Equation \eqref{eq:balance} states that, under the assumption of suitable boundary conditions, what is lost by the large scale energies either cascades to smaller scales or is transferred to thermal energy. In this view the pressure-strain interaction plays the role of an energy sink, reason for which we will refer to it, even if somehow inappropriately, as \emph{dissipation}.  \\

\noindent We evaluate equation \eqref{eq:balance} between scales $\ell_0$ and $\ell<\ell_0$. Under the hypotheses of negligible spatial fluxes and energy stationarity at scales smaller than $\ell_0$ we find:
\begin{equation}
    \Pi(\ell_0)-\Pi(\ell)=-\Delta\mathrm{PS}_i(\ell)-\Delta\mathrm{PS}_e(\ell)
    \label{eq:delta_balance}
\end{equation}
where $-\Delta\mathrm{PS}_\alpha(\ell)=-\mathrm{PS}_\alpha(\ell)+\mathrm{PS}_\alpha(\ell_0)$ is the cumulative contribution to thermal heating rate of species $\alpha$ in the range $[\ell,\ell_0]$.
\eqref{eq:delta_balance} shows that any difference between $\Pi(\ell_0)$ and $\Pi(\ell)$ indicates the amount of energy that is lost to thermal energy between those two scales. In this perspective a constant cascade rate indicates an \textit{inertial range} where dissipation is negligible, while a scale-dependent cascade rate is signature of active dissipation. It must be stressed that relation \eqref{eq:balance} comes directly from Vlasov-Maxwell equations, and as such is not limited by any fluid approximation since no closure equation is imposed on the pressure. This implies that if kinetic effects play a role in heating the plasma, this will be captured by the $\mathrm{PS}$ interaction, which explains why the energy cascade rate (inherently a fluid quantity) could capture dissipation via Landau damping in turbulence simulations \citep{Ferrand_2021LF}. However, the interpretation of $\mathrm{PS}$ as a measure of change in the thermal energy (i.e., heating) is grounded on the assumption that the spatial fluxes contribution (including the heat flux) in equation \eqref{eq_th} are negligible.

\noindent \textit{Data selection and methods--} We use data from the Magnetospheric Multispacecraft (MMS) mission \citep{Burch2016}, which enables us to compute the spatial derivatives in $\mathrm{PS}_\alpha$ and $\Pi$ using the gradiometer technique \citep{chanteur_spatial_1998}. We use FluxGate Magnetometer data for the magnetic field, the Spin-Plane Double Probe \citep{lindqvist_spin-plane_2016} and the Axial Double Probe \citep{ergun_axial_2016} for the electric field and the Fast Plasma Investigation \citep{pollock_fast_2016} for the plasma data. Spin-tone removal is applied to the electron velocity data.
The CG operation in spacecraft data at a given time-scale $\tau$ is computed as $\overline{f}_\tau(t)=\int dt' G_\tau(t') f(t+t')$ where $G_\tau$ is a 1-dimensional filtering kernel with characteristic width $\sim\tau$.
 To minimize the finite sample size effect, the maximum scale $\tau_0$ should be significantly smaller than the duration of the interval under considerations. Here we set $\tau_0\approx 20-30\,s$, which limits the largest scales of the study to  $k\rho_i\gtrsim 0.05-0.1$ depending on the value of the Taylor shifted Larmor radius. The smallest accessible scale $\tau_{\mathrm{min}}$ is given by the instrument resolution: when combining MMS products of different time resolution, we re-sample the data at the frequency of the least resolved quantities (typically the ions at 150 m$s$), and set $\tau_\mathrm{min}$ at twice the resolution of the instrument. It is worth noting that the electron contributions to the cascade rate and $\mathrm{PS}_e$ do not involve ion data, and as such they are computed down to (twice) the electron data time resolution of $30 \,\mathrm{m}s$.\\
To ensure the robustness of the results with respect to the choice of the start an final time of each interval we follow this pipeline: we compute $\pi,\overline{P_i}:\nabla\overline{v_i},\overline{P_e}:\nabla\overline{v_e}$ for all the data points in the interval $[t_\mathrm{start},t_\mathrm{end}]$. We then average the above quantities in an interval $[t_\mathrm{start}+\Delta t_1, t_\mathrm{end}-\Delta t_2]$. By varying independently $\Delta t_1,\Delta t_2$ between 0 and 10\% of the interval duration we obtain different estimates of the cascade rate $\Pi$ and $\mathrm{PS}_\alpha$. At each scale we take the mean value as our best estimate and use the standard deviation as error bar estimate. 

\noindent \textit{Results--} 
\begin{figure}
    \centering
    \includegraphics[width=0.5\textwidth]{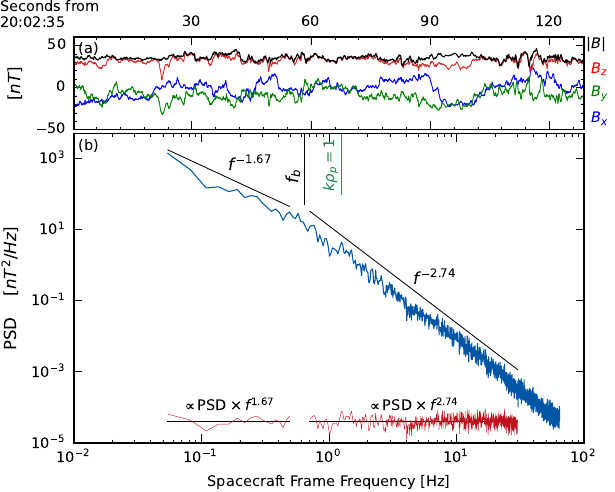}
    \caption{Panel (a) shows the time series of the magnetic field measured by MMS1 in Geocentric Solar Ecliptic (GSE) coordinates.
    Panel (b) displays the power spectrum of the magnetic field data computed using the Welch method \citep{Welch}. Power-law fit and compensated spectra are shown.}
    \label{fig:Spectrum}
\end{figure}
We show in \figref{fig:Spectrum} the data from MMS1 taken in the Earth's magnetosheath (2016/02/23, 20:02:35 - 20:04:44). During this time the average plasma conditions were: $B\approx 35\,$n$T$, $n_e\approx 19\, cm^{-3}$, $T_i \approx 175\,eV$ $T_e \approx 27\,eV$. The ratio of thermal to magnetic pressure is $\beta_i\approx 1.07, \beta_e\approx 0.16$. The mean flow speed $V_f\approx 300\,km/s$ and the angle between the flow and the magnetic field $\theta_{\mathbf{v}\mathbf{B}}\approx 50^o$. The inter-spacecraft separation is of $\sim 11\,$k$m$. The magnetic field power spectrum (Fig. \ref{fig:Spectrum}) displays a $f^{-5/3}$ scaling in the MHD range and steepens to $f^{-2.74}$ at higher frequencies.\\
 \noindent\figref{fig:Balance}(a) shows the balance between the energy cascade rate $\Pi$ and the dissipation rate as a function of scale $\tau$. At large scales, within the inertial range, the dominant process is the cascade, while at smaller scales, approaching $k\rho_i\sim 0.2 $, dissipation grows and the turbulent cascade is progressively weakened. In accordance with equation \eqref{eq:delta_balance} the decline in the cascade rate is counterbalanced by a rise in ion and electron's $\mathrm{PS}$, maintaining the sum of the three quantities constant over two decades of scales. Furthermore, \figref{fig:Balance}(a) shows that the small scale edge of the MHD range is highly dissipative as the cascade rate weakens by a factor $\sim 5$, consistent with the idea of increased dissipation around the spectral break \citep{Hellinger_2022}. Throughout the weakly dissipative subion range the cascade rate keeps weakening progressively.\\
 \begin{figure}
    \centering
    \includegraphics[width=0.5\textwidth]{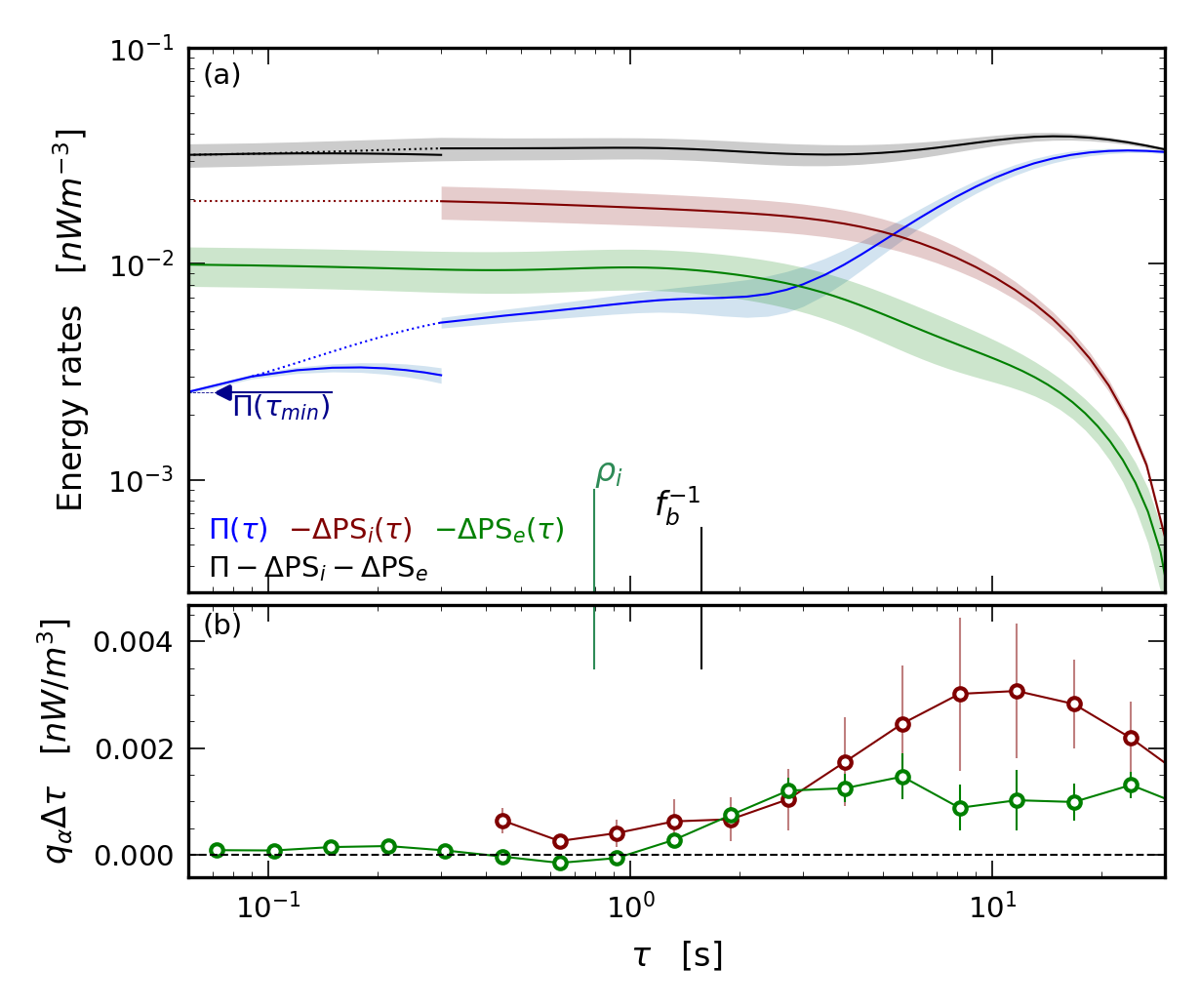}
    \caption{Panel (a) shows the different terms of equation \eqref{eq:delta_balance} as a function of scale for the interval shown in \figref{fig:Spectrum}. Shaded region denote the error bars. At time lags smaller than the time resolution of the ion data we assume that there is no additional contribution to ion heating (hence a constant $\Delta\mathrm{PS}_i(\tau)$, dotted), thus only the electron contribution to the cascade is computed. The (blue) dotted line is a cubic spline interpolation to aid the visualization. Panel (b) shows the scale-by-scale ion and electron heating rates (see text).}
    \label{fig:Balance}
\end{figure}
\noindent Looking at the behaviour of quantities $\Delta \mathrm{PS}_\alpha$ it is not immediate to infer at which scales ions and electrons are heated most. $\mathrm{PS}_\alpha$ being  a cumulative quantity, the contribution to the heating rate of a given scale range ($\tau$,$\tau+\Delta\tau$) is simply $q_\alpha(\tau)\Delta\tau\equiv-\mathrm{PS}_\alpha(\tau)+\mathrm{PS}_\alpha(\tau+\Delta\tau)\sim -[\partial \mathrm{PS}_\alpha/{\partial \tau}]\Delta\tau$, which is the quantity plotted in \figref{fig:Balance}(b) after binning logarithmically the range of scales.\\
\noindent The total heating rate for each species is defined as $Q_\alpha=\int_{\tau_\mathrm{min}}^{\tau_\mathrm{max}} q_\alpha(\tau)d\tau=\mathrm{PS}_\alpha(\tau_{\mathrm{max}})-\mathrm{PS}_\alpha(\tau_{\mathrm{min}})$, $\tau_\mathrm{min}=0.3\,s \;(0.06\,s)$ for ions (electrons) is twice the plasma data resolution and $\tau_\mathrm{max}=30\,s$ for this interval. For ions we obtain a total heating rate $Q_i=(1.9\pm0.3)\times 10^{-2}\,nW/m^3$ (assuming no additional contribution to ion heating originates from scales smaller than $0.3\,s$). For electrons, we estimate the heating rate $Q_e$ similarly to ions. However, that quantity is complemented by an extra term given by the energy cascade rate at the smallest available time lag, namely $\Pi(\tau=\tau_{min})$ (highlighted in \figref{fig:Balance}(a) for $\tau_{min}\sim 0.06$s), i.e. $Q_e^\star=Q_e+\Pi(\tau_\mathrm{min})$. This is based on the assumption that the residual cascade rate $\Pi(\tau_{min})$ will be entirely converted into electron heating at the smallest scales. Thus, we obtain the value $Q_e^\star=(1.2\pm 0.2)\times 10^{-2}\, nW/m^3$ for the total electron heating rate. The cascade rate at MHD scales has a value of $\Pi(\tau_\mathrm{max})=(3.04\pm 0.05)\times 10^{-2}\, nW/m^3$ placing the ratio $\Pi(\tau_\mathrm{max})/(Q_i+Q_e^\star)\simeq1$. We thus verify \eqref{eq:delta_balance} : what is lost by the energy cascade has been transferred to thermal energy. This observation confirms previous numerical results where the balance between the cascade rate and dissipation via $\mathrm{PS}$ was first reported \citep{Yang17,Hellinger_2022,Yang_2022} and improves over the comparison between $\mathrm{PS}$ and the cascade rate estimate via incompressible third order laws presented in \citep{Roy_2022}.

\noindent The study of the scale dependent heating rate $q_\alpha(\tau)\Delta\tau$ informs us about the scales most effective in heating the two species. Figure \ref{fig:Balance}(b) shows that the largest contribution to ion heating comes from $\tau\approx 10\,s$ (or $k\rho_i\approx 0.1$ using the Taylor hypothesis $\ell\sim \tau V_f\sim1/k$), in the same range of scales electrons are substantially heated. The residual cascade rate at $\tau_\mathrm{min}=0.06$,  assumed to sustain the turbulence and eventually heat electrons at scales $k\rho_i\gtrsim 13$ \citep{sahraoui2009}, accounts for $\Pi(\tau_\mathrm{min})/Q_e^\star\sim 30\%$  of the total electron heating rate.\\ 

\begin{figure}
    \centering
    \includegraphics[width=0.5\textwidth]{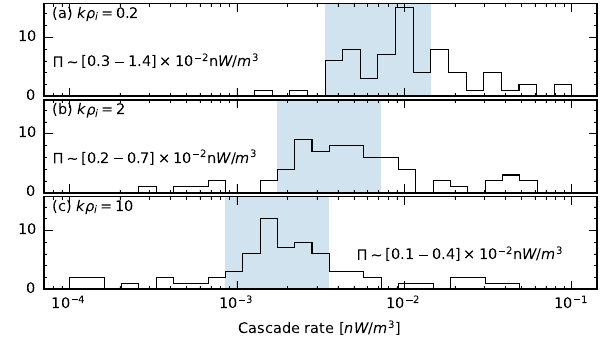}
    \caption{Histogram of the cascade rate at different scales, $k\rho_i=0.2, 2, 10$ respectively. The shaded blue region denotes the minimum number of contiguous bins that contains 60\% of the data set.}
    \label{fig:cascade_Rate}
\end{figure}

\noindent \textit{Statistics --} To confirm the statistical robustness of the previous results we perform the same analysis for a large sets of MMS intervals taken in the magnetosheath between October 2015 and May 2018 (we paid attention to exclude intervals that showed sharp jumps in the background plasma parameters). We further narrowed down the selection to keep only data intervals that satisfy the energy balance defined (within the error bars) by $\Pi(\tau_\mathrm{max})/(Q_i+ Q_e^\star) \in [0.4,1.6]$ to ensures that spatial fluxes and time derivatives in equation \eqref{eq:balance} are negligible in the range of scales we are considering. This guarantees a reliable estimation of the cascade rate, the ion and electron heating  rates and the related effective dissipation scales. To increase the size of the statistical sample, for a given interval we consider each spacecraft as an independent realization, although the spatial derivatives involved in equation \eqref{eq:delta_balance} are computed from the four spacecraft and thus are identical for a given event. \\
\begin{figure}
    \centering
    \includegraphics[width=0.5\textwidth]{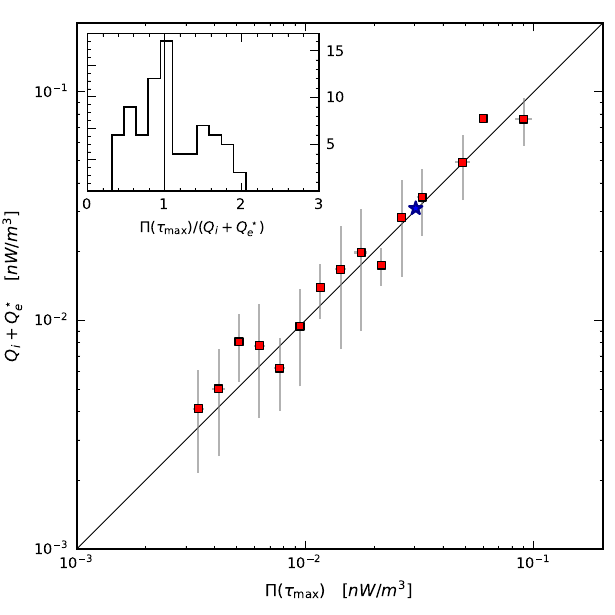}
    \caption{Cascade-Dissipation balance for the selected intervals that satisfy the balance condition (see text) binned according to the value of the cascade rate $\Pi(\tau_{\mathrm{max}})$. The star denotes the case study presented in the text. The inset shows the histogram of the cascade-dissipation ratio for all the inervals.}
    \label{fig:stats_balance}
\end{figure}
\noindent Fig. \ref{fig:stats_balance} summarizes the cascade-dissipation balance for the selected intervals, which we use to estimate the value of the energy cascade rate in the Earth magnetosheath at different scales: the small scale edge of the MHD range $k\rho_i=0.2$, around the ion scale $k\rho_i=2$ and at the subion scale $k\rho_i=10$. Histograms of the cascade rate at different scales are displayed in \figref{fig:cascade_Rate}. To provide a statistically significant measure we identify the minimum number of consecutive logarithmically spaced bins containing over 60\% of the data set.  
Figure \ref{fig:cascade_Rate}(a) shows that MHD scales generally exhibit a cascade rate in the interval $[0.3-1.4]\times10^{-2}\, \mathrm{n}W/m^3$, comparable with the values reported in \citep{hadid18,andres2019}  using third order laws in the Earth's magnetosheath. 
Notably, this rate diminishes by a factor of two upon reaching $k\rho_i=2$ [\figref{fig:cascade_Rate}(b)] and further weakens by an additional factor of two at  $k\rho_i=10$ [\figref{fig:cascade_Rate}(c)] reaching a rate in the range $[0.1-0.4]\times 10^{-2} \mathrm{n}W/m^3$. This shows that the subion range is weakly dissipative and that at $k\rho_i\sim 10$ a significant ratio ($\sim 30\%$) of the cascade rate at MHD scales remains available to sustain the turbulence cascade all the way down to electron scales \citep{sahraoui2009}.

\begin{figure}
    \centering
    \includegraphics[width=0.5\textwidth]{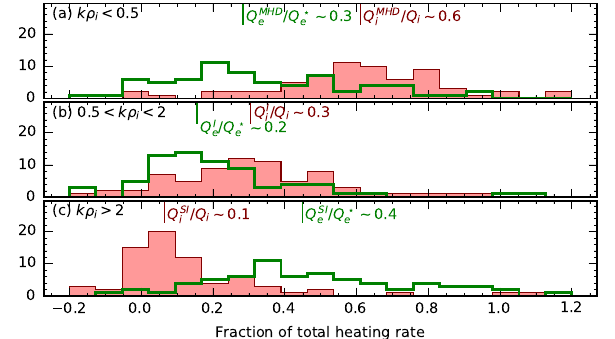}
    \caption{ Histogram of the relative contribution to the heating rate from the three range of scales $k\rho_i<0.5$ (a), $0.5<k\rho_i<2$ (b) $k\rho_i>2$ Ions (red,filled) and electrons (green, empty). In each panel a vertical sign denotes the median for each populatuion.  }
    \label{fig:hist_heating}
\end{figure}

We now wish to delineate the scales at which the $\mathrm{PS}$ interaction is effective in heating the plasma. For each species we calculate the fraction of heating coming from the MHD range ($k\rho_i<0.5$), around the ion Larmor scale ($0.5<k\rho_i<2$) and the subion range ($k\rho_i>2$).  
The results in \figref{fig:hist_heating}(a-c) show that the largest contribution to the ion heating rate comes from MHD scales (the median contribution being 60\%). The relative importance decreases to 30\% at the ion Larmor scale and then 10\% at sub-Larmor scales. This result corroborates the assumption made above that the residual cascade at $k\rho_i\gtrsim 10$ translates predominantly into electron heating. \\ 
The picture that emerges for electrons is more complex: \figref{fig:hist_heating}(a-c) show a nearly equal contribution from all scale ranges: 30\%, 20\% and 40\% median contribution from the MHD, ion Larmor and subion scales, respectively. This result demonstrates that electron heating can be significant at scales comparable with the ion Larmor radius (including the edge of the MHD range) in line with some numerical results \citep{howes2011}.
\noindent \textit{Conclusions --}
In this work we measure for the first time using in-situ data the scale dependence of the cascade rate {\it and} the dissipation rate and show that there exist a balance that holds for over two decades of scales: the depletion of the energy cascade as turbulence proceeds from MHD to kinetic scales is compensated by a net positive transfer to the thermal energy. On a statistical data set we show that electrons can get substantial heating at scales comparable with the ion Larmor radius, against the conventional wisdom that electron heating occurs solely at electron scales. This questions the validity of kinetic-hybrid models for plasmas where electrons are treated as a politropic fluid whose dissipation is confined to the small (electron) resistive scales. \\
\noindent Despite the net decline of the cascade rate in the subion range, the magnetic energy spectra still show clear power-laws all the way down to the electron scales. In the absence of a rigorous explanation to this observation, we speculate that any residual turbulent energy is bound to cascade to small scales following the scaling law of one of the existing modes (e.g., Kinetic Alfv\'en Modes \citep{sahraoui2010,howes2011}. \\
\noindent The technique used in this work enables us to quantify turbulent heating rate, but it does not inform us about the processes responsible for it. In this view methods based on the analysis of the velocity distribution functions \citep{howes_klein_li_2017,schekochihinHermite,servidioHermite,cassak_2023} should be seen as complementary to this study as they can provide the missing part of the information about the nature of the processes at play.

\begin{acknowledgements}
\noindent DM acknowledges useful discussions with A. Chasapis. MMS data come from CDPP/AMDA and NASA GSFC's Space Physics Data Facility's CDAWeb. The python client SPEASY was used for data retrieval.
\end{acknowledgements}

\bibliographystyle{apsrev4-1} 

\begin{thebibliography}{53}%
\makeatletter
\providecommand \@ifxundefined [1]{%
 \@ifx{#1\undefined}
}%
\providecommand \@ifnum [1]{%
 \ifnum #1\expandafter \@firstoftwo
 \else \expandafter \@secondoftwo
 \fi
}%
\providecommand \@ifx [1]{%
 \ifx #1\expandafter \@firstoftwo
 \else \expandafter \@secondoftwo
 \fi
}%
\providecommand \natexlab [1]{#1}%
\providecommand \enquote  [1]{``#1''}%
\providecommand \bibnamefont  [1]{#1}%
\providecommand \bibfnamefont [1]{#1}%
\providecommand \citenamefont [1]{#1}%
\providecommand \href@noop [0]{\@secondoftwo}%
\providecommand \href [0]{\begingroup \@sanitize@url \@href}%
\providecommand \@href[1]{\@@startlink{#1}\@@href}%
\providecommand \@@href[1]{\endgroup#1\@@endlink}%
\providecommand \@sanitize@url [0]{\catcode `\\12\catcode `\$12\catcode
  `\&12\catcode `\#12\catcode `\^12\catcode `\_12\catcode `\%12\relax}%
\providecommand \@@startlink[1]{}%
\providecommand \@@endlink[0]{}%
\providecommand \url  [0]{\begingroup\@sanitize@url \@url }%
\providecommand \@url [1]{\endgroup\@href {#1}{\urlprefix }}%
\providecommand \urlprefix  [0]{URL }%
\providecommand \Eprint [0]{\href }%
\providecommand \doibase [0]{http://dx.doi.org/}%
\providecommand \selectlanguage [0]{\@gobble}%
\providecommand \bibinfo  [0]{\@secondoftwo}%
\providecommand \bibfield  [0]{\@secondoftwo}%
\providecommand \translation [1]{[#1]}%
\providecommand \BibitemOpen [0]{}%
\providecommand \bibitemStop [0]{}%
\providecommand \bibitemNoStop [0]{.\EOS\space}%
\providecommand \EOS [0]{\spacefactor3000\relax}%
\providecommand \BibitemShut  [1]{\csname bibitem#1\endcsname}%
\let\auto@bib@innerbib\@empty
\bibitem [{\citenamefont {Politano}\ and\ \citenamefont
  {Pouquet}(1998)}]{politano1998}%
  \BibitemOpen
  \bibfield  {author} {\bibinfo {author} {\bibfnamefont {H.}~\bibnamefont
  {Politano}}\ and\ \bibinfo {author} {\bibfnamefont {A.}~\bibnamefont
  {Pouquet}},\ }\href {\doibase 10.1103/PhysRevE.57.R21} {\bibfield  {journal}
  {\bibinfo  {journal} {Phys. Rev. E}\ }\textbf {\bibinfo {volume} {57}},\
  \bibinfo {pages} {R21} (\bibinfo {year} {1998})}\BibitemShut {NoStop}%
\bibitem [{\citenamefont {Sahraoui}\ \emph {et~al.}(2020)\citenamefont
  {Sahraoui}, \citenamefont {Hadid},\ and\ \citenamefont
  {Huang}}]{sahraoui2020}%
  \BibitemOpen
  \bibfield  {author} {\bibinfo {author} {\bibfnamefont {F.}~\bibnamefont
  {Sahraoui}}, \bibinfo {author} {\bibfnamefont {L.}~\bibnamefont {Hadid}}, \
  and\ \bibinfo {author} {\bibfnamefont {S.}~\bibnamefont {Huang}},\ }\href
  {\doibase 10.1007/s41614-020-0040-2} {\bibfield  {journal} {\bibinfo
  {journal} {Rev. Mod. Phys.}\ }\textbf {\bibinfo {volume} {4}},\ \bibinfo
  {pages} {4} (\bibinfo {year} {2020})}\BibitemShut {NoStop}%
\bibitem [{\citenamefont {Banerjee}\ and\ \citenamefont
  {Galtier}(2013)}]{banerjee2013}%
  \BibitemOpen
  \bibfield  {author} {\bibinfo {author} {\bibfnamefont {S.}~\bibnamefont
  {Banerjee}}\ and\ \bibinfo {author} {\bibfnamefont {S.}~\bibnamefont
  {Galtier}},\ }\href {\doibase 10.1103/PhysRevE.87.013019} {\bibfield
  {journal} {\bibinfo  {journal} {Phys. Rev. E}\ }\textbf {\bibinfo {volume}
  {87}},\ \bibinfo {pages} {013019} (\bibinfo {year} {2013})}\BibitemShut
  {NoStop}%
\bibitem [{\citenamefont {Andrés}\ \emph {et~al.}(2018)\citenamefont
  {Andrés}, \citenamefont {Galtier},\ and\ \citenamefont
  {Sahraoui}}]{andres2018}%
  \BibitemOpen
  \bibfield  {author} {\bibinfo {author} {\bibfnamefont {N.}~\bibnamefont
  {Andrés}}, \bibinfo {author} {\bibfnamefont {S.}~\bibnamefont {Galtier}}, \
  and\ \bibinfo {author} {\bibfnamefont {F.}~\bibnamefont {Sahraoui}},\ }\href
  {\doibase 10.1103/PhysRevE.97.013204} {\bibfield  {journal} {\bibinfo
  {journal} {Phys. Rev. E}\ }\textbf {\bibinfo {volume} {97}},\ \bibinfo
  {pages} {013204} (\bibinfo {year} {2018})}\BibitemShut {NoStop}%
\bibitem [{\citenamefont {Galtier}(2008)}]{galtier2008}%
  \BibitemOpen
  \bibfield  {author} {\bibinfo {author} {\bibfnamefont {S.}~\bibnamefont
  {Galtier}},\ }\href {\doibase 10.1103/PhysRevE.77.015302} {\bibfield
  {journal} {\bibinfo  {journal} {Phys. Rev. E}\ }\textbf {\bibinfo {volume}
  {77}},\ \bibinfo {pages} {015302} (\bibinfo {year} {2008})}\BibitemShut
  {NoStop}%
\bibitem [{\citenamefont {Hellinger}\ \emph {et~al.}(2018)\citenamefont
  {Hellinger}, \citenamefont {Verdini}, \citenamefont {Landi}, \citenamefont
  {Franci},\ and\ \citenamefont {Matteini}}]{hellinger2018}%
  \BibitemOpen
  \bibfield  {author} {\bibinfo {author} {\bibfnamefont {P.}~\bibnamefont
  {Hellinger}}, \bibinfo {author} {\bibfnamefont {A.}~\bibnamefont {Verdini}},
  \bibinfo {author} {\bibfnamefont {S.}~\bibnamefont {Landi}}, \bibinfo
  {author} {\bibfnamefont {L.}~\bibnamefont {Franci}}, \ and\ \bibinfo {author}
  {\bibfnamefont {L.}~\bibnamefont {Matteini}},\ }\href {\doibase
  10.3847/2041-8213/aabc06} {\bibfield  {journal} {\bibinfo  {journal}
  {Astrophys. J.(Lett.)}\ }\textbf {\bibinfo {volume} {857}},\ \bibinfo {pages}
  {L19} (\bibinfo {year} {2018})}\BibitemShut {NoStop}%
\bibitem [{\citenamefont {Ferrand}\ \emph {et~al.}(2019)\citenamefont
  {Ferrand}, \citenamefont {Galtier}, \citenamefont {Sahraoui}, \citenamefont
  {Meyrand}, \citenamefont {Andrés},\ and\ \citenamefont
  {Banerjee}}]{ferrand19}%
  \BibitemOpen
  \bibfield  {author} {\bibinfo {author} {\bibfnamefont {R.}~\bibnamefont
  {Ferrand}}, \bibinfo {author} {\bibfnamefont {S.}~\bibnamefont {Galtier}},
  \bibinfo {author} {\bibfnamefont {F.}~\bibnamefont {Sahraoui}}, \bibinfo
  {author} {\bibfnamefont {R.}~\bibnamefont {Meyrand}}, \bibinfo {author}
  {\bibfnamefont {N.}~\bibnamefont {Andrés}}, \ and\ \bibinfo {author}
  {\bibfnamefont {S.}~\bibnamefont {Banerjee}},\ }\href {\doibase
  10.3847/1538-4357/ab2be9} {\bibfield  {journal} {\bibinfo  {journal}
  {Astrophys. J.}\ }\textbf {\bibinfo {volume} {881}},\ \bibinfo {pages} {50}
  (\bibinfo {year} {2019})}\BibitemShut {NoStop}%
\bibitem [{\citenamefont {Simon}\ and\ \citenamefont
  {Sahraoui}(2021)}]{simon2021}%
  \BibitemOpen
  \bibfield  {author} {\bibinfo {author} {\bibfnamefont {P.}~\bibnamefont
  {Simon}}\ and\ \bibinfo {author} {\bibfnamefont {F.}~\bibnamefont
  {Sahraoui}},\ }\href {\doibase 10.3847/1538-4357/ac0337} {\bibfield
  {journal} {\bibinfo  {journal} {Astrophys. J.}\ }\textbf {\bibinfo {volume}
  {916}},\ \bibinfo {pages} {49} (\bibinfo {year} {2021})}\BibitemShut
  {NoStop}%
\bibitem [{\citenamefont {Simon}\ and\ \citenamefont
  {Sahraoui}(2022)}]{simon2022}%
  \BibitemOpen
  \bibfield  {author} {\bibinfo {author} {\bibfnamefont {P.}~\bibnamefont
  {Simon}}\ and\ \bibinfo {author} {\bibfnamefont {F.}~\bibnamefont
  {Sahraoui}},\ }\href {\doibase 10.1103/PhysRevE.105.055111} {\bibfield
  {journal} {\bibinfo  {journal} {Phys. Rev. E}\ }\textbf {\bibinfo {volume}
  {105}},\ \bibinfo {pages} {055111} (\bibinfo {year} {2022})}\BibitemShut
  {NoStop}%
\bibitem [{\citenamefont {Sorriso-Valvo}\ \emph {et~al.}(2007)\citenamefont
  {Sorriso-Valvo}, \citenamefont {Marino}, \citenamefont {Carbone},
  \citenamefont {Noullez}, \citenamefont {Lepreti}, \citenamefont {Veltri},
  \citenamefont {Bruno}, \citenamefont {Bavassano},\ and\ \citenamefont
  {Pietropaolo}}]{sorriso-valvo2007}%
  \BibitemOpen
  \bibfield  {author} {\bibinfo {author} {\bibfnamefont {L.}~\bibnamefont
  {Sorriso-Valvo}}, \bibinfo {author} {\bibfnamefont {R.}~\bibnamefont
  {Marino}}, \bibinfo {author} {\bibfnamefont {V.}~\bibnamefont {Carbone}},
  \bibinfo {author} {\bibfnamefont {A.}~\bibnamefont {Noullez}}, \bibinfo
  {author} {\bibfnamefont {F.}~\bibnamefont {Lepreti}}, \bibinfo {author}
  {\bibfnamefont {P.}~\bibnamefont {Veltri}}, \bibinfo {author} {\bibfnamefont
  {R.}~\bibnamefont {Bruno}}, \bibinfo {author} {\bibfnamefont
  {B.}~\bibnamefont {Bavassano}}, \ and\ \bibinfo {author} {\bibfnamefont
  {E.}~\bibnamefont {Pietropaolo}},\ }\href {\doibase
  10.1103/PhysRevLett.99.115001} {\bibfield  {journal} {\bibinfo  {journal}
  {Phys. Rev. Lett.}\ }\textbf {\bibinfo {volume} {99}},\ \bibinfo {pages}
  {115001} (\bibinfo {year} {2007})}\BibitemShut {NoStop}%
\bibitem [{\citenamefont {MacBride}\ \emph {et~al.}(2008)\citenamefont
  {MacBride}, \citenamefont {Smith},\ and\ \citenamefont
  {Forman}}]{MacBride_2008}%
  \BibitemOpen
  \bibfield  {author} {\bibinfo {author} {\bibfnamefont {B.~T.}\ \bibnamefont
  {MacBride}}, \bibinfo {author} {\bibfnamefont {C.~W.}\ \bibnamefont {Smith}},
  \ and\ \bibinfo {author} {\bibfnamefont {M.~A.}\ \bibnamefont {Forman}},\
  }\href {\doibase 10.1086/529575} {\bibfield  {journal} {\bibinfo  {journal}
  {Astrophys. J.}\ }\textbf {\bibinfo {volume} {679}},\ \bibinfo {pages} {1644}
  (\bibinfo {year} {2008})}\BibitemShut {NoStop}%
\bibitem [{\citenamefont {Stawarz}\ \emph {et~al.}(2009)\citenamefont
  {Stawarz}, \citenamefont {Smith}, \citenamefont {Vasquez}, \citenamefont
  {Forman},\ and\ \citenamefont {MacBride}}]{Stawarz_2009}%
  \BibitemOpen
  \bibfield  {author} {\bibinfo {author} {\bibfnamefont {J.~E.}\ \bibnamefont
  {Stawarz}}, \bibinfo {author} {\bibfnamefont {C.~W.}\ \bibnamefont {Smith}},
  \bibinfo {author} {\bibfnamefont {B.~J.}\ \bibnamefont {Vasquez}}, \bibinfo
  {author} {\bibfnamefont {M.~A.}\ \bibnamefont {Forman}}, \ and\ \bibinfo
  {author} {\bibfnamefont {B.~T.}\ \bibnamefont {MacBride}},\ }\href {\doibase
  10.1088/0004-637X/697/2/1119} {\bibfield  {journal} {\bibinfo  {journal}
  {Astrophys. J.}\ }\textbf {\bibinfo {volume} {697}},\ \bibinfo {pages} {1119}
  (\bibinfo {year} {2009})}\BibitemShut {NoStop}%
\bibitem [{\citenamefont {Stawarz}\ \emph {et~al.}(2010)\citenamefont
  {Stawarz}, \citenamefont {Smith}, \citenamefont {Vasquez}, \citenamefont
  {Forman},\ and\ \citenamefont {MacBride}}]{stawarz10}%
  \BibitemOpen
  \bibfield  {author} {\bibinfo {author} {\bibfnamefont {J.~E.}\ \bibnamefont
  {Stawarz}}, \bibinfo {author} {\bibfnamefont {C.~W.}\ \bibnamefont {Smith}},
  \bibinfo {author} {\bibfnamefont {B.~J.}\ \bibnamefont {Vasquez}}, \bibinfo
  {author} {\bibfnamefont {M.~A.}\ \bibnamefont {Forman}}, \ and\ \bibinfo
  {author} {\bibfnamefont {B.~T.}\ \bibnamefont {MacBride}},\ }\href {\doibase
  10.1088/0004-637X/713/2/920} {\bibfield  {journal} {\bibinfo  {journal}
  {Astrophys. J.}\ }\textbf {\bibinfo {volume} {713}},\ \bibinfo {pages} {920}
  (\bibinfo {year} {2010})}\BibitemShut {NoStop}%
\bibitem [{\citenamefont {Coburn}\ \emph {et~al.}(2012)\citenamefont {Coburn},
  \citenamefont {Smith}, \citenamefont {Vasquez}, \citenamefont {Stawarz},\
  and\ \citenamefont {Forman}}]{Coburn_2012}%
  \BibitemOpen
  \bibfield  {author} {\bibinfo {author} {\bibfnamefont {J.~T.}\ \bibnamefont
  {Coburn}}, \bibinfo {author} {\bibfnamefont {C.~W.}\ \bibnamefont {Smith}},
  \bibinfo {author} {\bibfnamefont {B.~J.}\ \bibnamefont {Vasquez}}, \bibinfo
  {author} {\bibfnamefont {J.~E.}\ \bibnamefont {Stawarz}}, \ and\ \bibinfo
  {author} {\bibfnamefont {M.~A.}\ \bibnamefont {Forman}},\ }\href {\doibase
  10.1088/0004-637X/754/2/93} {\bibfield  {journal} {\bibinfo  {journal}
  {Astrophys. J.}\ }\textbf {\bibinfo {volume} {754}},\ \bibinfo {pages} {93}
  (\bibinfo {year} {2012})}\BibitemShut {NoStop}%
\bibitem [{\citenamefont {Hadid}\ \emph {et~al.}(2017)\citenamefont {Hadid},
  \citenamefont {Sahraoui},\ and\ \citenamefont {Galtier}}]{hadid17}%
  \BibitemOpen
  \bibfield  {author} {\bibinfo {author} {\bibfnamefont {L.~Z.}\ \bibnamefont
  {Hadid}}, \bibinfo {author} {\bibfnamefont {F.}~\bibnamefont {Sahraoui}}, \
  and\ \bibinfo {author} {\bibfnamefont {S.}~\bibnamefont {Galtier}},\ }\href
  {\doibase 10.3847/1538-4357/aa603f} {\bibfield  {journal} {\bibinfo
  {journal} {Astrophys. J.}\ }\textbf {\bibinfo {volume} {838}},\ \bibinfo
  {pages} {9} (\bibinfo {year} {2017})}\BibitemShut {NoStop}%
\bibitem [{\citenamefont {Hadid}\ \emph {et~al.}(2018)\citenamefont {Hadid},
  \citenamefont {Sahraoui}, \citenamefont {Galtier},\ and\ \citenamefont
  {Huang}}]{hadid18}%
  \BibitemOpen
  \bibfield  {author} {\bibinfo {author} {\bibfnamefont {L.~Z.}\ \bibnamefont
  {Hadid}}, \bibinfo {author} {\bibfnamefont {F.}~\bibnamefont {Sahraoui}},
  \bibinfo {author} {\bibfnamefont {S.}~\bibnamefont {Galtier}}, \ and\
  \bibinfo {author} {\bibfnamefont {S.~Y.}\ \bibnamefont {Huang}},\ }\href
  {\doibase 10.1103/PhysRevLett.120.055102} {\bibfield  {journal} {\bibinfo
  {journal} {Phys. Rev. Lett.}\ }\textbf {\bibinfo {volume} {120}},\ \bibinfo
  {pages} {055102} (\bibinfo {year} {2018})}\BibitemShut {NoStop}%
\bibitem [{\citenamefont {Bandyopadhyay}\ \emph {et~al.}(2018)\citenamefont
  {Bandyopadhyay}, \citenamefont {Chasapis}, \citenamefont {Chhiber},
  \citenamefont {Parashar}, \citenamefont {Matthaeus}, \citenamefont {Shay},
  \citenamefont {Maruca}, \citenamefont {Burch}, \citenamefont {Moore},
  \citenamefont {Pollock}, \citenamefont {Giles}, \citenamefont {Paterson},
  \citenamefont {Dorelli}, \citenamefont {Gershman}, \citenamefont {Torbert},
  \citenamefont {Russell},\ and\ \citenamefont
  {Strangeway}}]{Bandyopadhyay_2018}%
  \BibitemOpen
  \bibfield  {author} {\bibinfo {author} {\bibfnamefont {R.}~\bibnamefont
  {Bandyopadhyay}}, \bibinfo {author} {\bibfnamefont {A.}~\bibnamefont
  {Chasapis}}, \bibinfo {author} {\bibfnamefont {R.}~\bibnamefont {Chhiber}},
  \bibinfo {author} {\bibfnamefont {T.~N.}\ \bibnamefont {Parashar}}, \bibinfo
  {author} {\bibfnamefont {W.~H.}\ \bibnamefont {Matthaeus}}, \bibinfo {author}
  {\bibfnamefont {M.~A.}\ \bibnamefont {Shay}}, \bibinfo {author}
  {\bibfnamefont {B.~A.}\ \bibnamefont {Maruca}}, \bibinfo {author}
  {\bibfnamefont {J.~L.}\ \bibnamefont {Burch}}, \bibinfo {author}
  {\bibfnamefont {T.~E.}\ \bibnamefont {Moore}}, \bibinfo {author}
  {\bibfnamefont {C.~J.}\ \bibnamefont {Pollock}}, \bibinfo {author}
  {\bibfnamefont {B.~L.}\ \bibnamefont {Giles}}, \bibinfo {author}
  {\bibfnamefont {W.~R.}\ \bibnamefont {Paterson}}, \bibinfo {author}
  {\bibfnamefont {J.}~\bibnamefont {Dorelli}}, \bibinfo {author} {\bibfnamefont
  {D.~J.}\ \bibnamefont {Gershman}}, \bibinfo {author} {\bibfnamefont {R.~B.}\
  \bibnamefont {Torbert}}, \bibinfo {author} {\bibfnamefont {C.~T.}\
  \bibnamefont {Russell}}, \ and\ \bibinfo {author} {\bibfnamefont {R.~J.}\
  \bibnamefont {Strangeway}},\ }\href {\doibase 10.3847/1538-4357/aade04}
  {\bibfield  {journal} {\bibinfo  {journal} {Astrophys. J.}\ }\textbf
  {\bibinfo {volume} {866}},\ \bibinfo {pages} {106} (\bibinfo {year}
  {2018})}\BibitemShut {NoStop}%
\bibitem [{\citenamefont {Andr\'es}\ \emph {et~al.}(2019)\citenamefont
  {Andr\'es}, \citenamefont {Sahraoui}, \citenamefont {Galtier}, \citenamefont
  {Hadid}, \citenamefont {Ferrand},\ and\ \citenamefont {Huang}}]{andres2019}%
  \BibitemOpen
  \bibfield  {author} {\bibinfo {author} {\bibfnamefont {N.}~\bibnamefont
  {Andr\'es}}, \bibinfo {author} {\bibfnamefont {F.}~\bibnamefont {Sahraoui}},
  \bibinfo {author} {\bibfnamefont {S.}~\bibnamefont {Galtier}}, \bibinfo
  {author} {\bibfnamefont {L.~Z.}\ \bibnamefont {Hadid}}, \bibinfo {author}
  {\bibfnamefont {R.}~\bibnamefont {Ferrand}}, \ and\ \bibinfo {author}
  {\bibfnamefont {S.~Y.}\ \bibnamefont {Huang}},\ }\href {\doibase
  10.1103/PhysRevLett.123.245101} {\bibfield  {journal} {\bibinfo  {journal}
  {Phys. Rev. Lett.}\ }\textbf {\bibinfo {volume} {123}},\ \bibinfo {pages}
  {245101} (\bibinfo {year} {2019})}\BibitemShut {NoStop}%
\bibitem [{\citenamefont {Bandyopadhyay}\ \emph {et~al.}(2020)\citenamefont
  {Bandyopadhyay}, \citenamefont {Sorriso-Valvo}, \citenamefont {Chasapis},
  \citenamefont {Hellinger}, \citenamefont {Matthaeus}, \citenamefont
  {Verdini}, \citenamefont {Landi}, \citenamefont {Franci}, \citenamefont
  {Matteini}, \citenamefont {Giles}, \citenamefont {Gershman}, \citenamefont
  {Moore}, \citenamefont {Pollock}, \citenamefont {Russell}, \citenamefont
  {Strangeway}, \citenamefont {Torbert},\ and\ \citenamefont
  {Burch}}]{Bandyopadhyay_2020}%
  \BibitemOpen
  \bibfield  {author} {\bibinfo {author} {\bibfnamefont {R.}~\bibnamefont
  {Bandyopadhyay}}, \bibinfo {author} {\bibfnamefont {L.}~\bibnamefont
  {Sorriso-Valvo}}, \bibinfo {author} {\bibfnamefont {A.}~\bibnamefont
  {Chasapis}}, \bibinfo {author} {\bibfnamefont {P.}~\bibnamefont {Hellinger}},
  \bibinfo {author} {\bibfnamefont {W.~H.}\ \bibnamefont {Matthaeus}}, \bibinfo
  {author} {\bibfnamefont {A.}~\bibnamefont {Verdini}}, \bibinfo {author}
  {\bibfnamefont {S.}~\bibnamefont {Landi}}, \bibinfo {author} {\bibfnamefont
  {L.}~\bibnamefont {Franci}}, \bibinfo {author} {\bibfnamefont
  {L.}~\bibnamefont {Matteini}}, \bibinfo {author} {\bibfnamefont {B.~L.}\
  \bibnamefont {Giles}}, \bibinfo {author} {\bibfnamefont {D.~J.}\ \bibnamefont
  {Gershman}}, \bibinfo {author} {\bibfnamefont {T.~E.}\ \bibnamefont {Moore}},
  \bibinfo {author} {\bibfnamefont {C.~J.}\ \bibnamefont {Pollock}}, \bibinfo
  {author} {\bibfnamefont {C.~T.}\ \bibnamefont {Russell}}, \bibinfo {author}
  {\bibfnamefont {R.~J.}\ \bibnamefont {Strangeway}}, \bibinfo {author}
  {\bibfnamefont {R.~B.}\ \bibnamefont {Torbert}}, \ and\ \bibinfo {author}
  {\bibfnamefont {J.~L.}\ \bibnamefont {Burch}},\ }\href {\doibase
  10.1103/PhysRevLett.124.225101} {\bibfield  {journal} {\bibinfo  {journal}
  {Phys. Rev. Lett.}\ }\textbf {\bibinfo {volume} {124}},\ \bibinfo {pages}
  {225101} (\bibinfo {year} {2020})}\BibitemShut {NoStop}%
\bibitem [{\citenamefont {Andrés}\ \emph {et~al.}(2021)\citenamefont
  {Andrés}, \citenamefont {Sahraoui}, \citenamefont {Hadid}, \citenamefont
  {Huang}, \citenamefont {Romanelli}, \citenamefont {Galtier}, \citenamefont
  {DiBraccio},\ and\ \citenamefont {Halekas}}]{andres_2021}%
  \BibitemOpen
  \bibfield  {author} {\bibinfo {author} {\bibfnamefont {N.}~\bibnamefont
  {Andrés}}, \bibinfo {author} {\bibfnamefont {F.}~\bibnamefont {Sahraoui}},
  \bibinfo {author} {\bibfnamefont {L.~Z.}\ \bibnamefont {Hadid}}, \bibinfo
  {author} {\bibfnamefont {S.~Y.}\ \bibnamefont {Huang}}, \bibinfo {author}
  {\bibfnamefont {N.}~\bibnamefont {Romanelli}}, \bibinfo {author}
  {\bibfnamefont {S.}~\bibnamefont {Galtier}}, \bibinfo {author} {\bibfnamefont
  {G.}~\bibnamefont {DiBraccio}}, \ and\ \bibinfo {author} {\bibfnamefont
  {J.}~\bibnamefont {Halekas}},\ }\href {\doibase 10.3847/1538-4357/ac0af5}
  {\bibfield  {journal} {\bibinfo  {journal} {Astrophys. J.}\ }\textbf
  {\bibinfo {volume} {919}},\ \bibinfo {pages} {19} (\bibinfo {year}
  {2021})}\BibitemShut {NoStop}%
\bibitem [{\citenamefont {Brodiano}\ \emph {et~al.}(2023)\citenamefont
  {Brodiano}, \citenamefont {Dmitruk},\ and\ \citenamefont
  {Andrés}}]{brodiano23}%
  \BibitemOpen
  \bibfield  {author} {\bibinfo {author} {\bibfnamefont {M.}~\bibnamefont
  {Brodiano}}, \bibinfo {author} {\bibfnamefont {P.}~\bibnamefont {Dmitruk}}, \
  and\ \bibinfo {author} {\bibfnamefont {N.}~\bibnamefont {Andrés}},\ }\href
  {\doibase 10.1063/5.0109379} {\bibfield  {journal} {\bibinfo  {journal}
  {Phys. Plasmas}\ }\textbf {\bibinfo {volume} {30}},\ \bibinfo {pages}
  {032903} (\bibinfo {year} {2023})}\BibitemShut {NoStop}%
\bibitem [{\citenamefont {Pecora}\ \emph {et~al.}(2023)\citenamefont {Pecora},
  \citenamefont {Yang}, \citenamefont {Matthaeus}, \citenamefont {Chasapis},
  \citenamefont {Klein}, \citenamefont {Stevens}, \citenamefont {Servidio},
  \citenamefont {Greco}, \citenamefont {Gershman}, \citenamefont {Giles},\ and\
  \citenamefont {Burch}}]{pecora23}%
  \BibitemOpen
  \bibfield  {author} {\bibinfo {author} {\bibfnamefont {F.}~\bibnamefont
  {Pecora}}, \bibinfo {author} {\bibfnamefont {Y.}~\bibnamefont {Yang}},
  \bibinfo {author} {\bibfnamefont {W.~H.}\ \bibnamefont {Matthaeus}}, \bibinfo
  {author} {\bibfnamefont {A.}~\bibnamefont {Chasapis}}, \bibinfo {author}
  {\bibfnamefont {K.~G.}\ \bibnamefont {Klein}}, \bibinfo {author}
  {\bibfnamefont {M.}~\bibnamefont {Stevens}}, \bibinfo {author} {\bibfnamefont
  {S.}~\bibnamefont {Servidio}}, \bibinfo {author} {\bibfnamefont
  {A.}~\bibnamefont {Greco}}, \bibinfo {author} {\bibfnamefont {D.~J.}\
  \bibnamefont {Gershman}}, \bibinfo {author} {\bibfnamefont {B.~L.}\
  \bibnamefont {Giles}}, \ and\ \bibinfo {author} {\bibfnamefont {J.~L.}\
  \bibnamefont {Burch}},\ }\href {\doibase 10.1103/PhysRevLett.131.225201}
  {\bibfield  {journal} {\bibinfo  {journal} {Phys. Rev. Lett.}\ }\textbf
  {\bibinfo {volume} {131}},\ \bibinfo {pages} {225201} (\bibinfo {year}
  {2023})}\BibitemShut {NoStop}%
\bibitem [{\citenamefont {Germano}(1992)}]{germano_turbulence_1992}%
  \BibitemOpen
  \bibfield  {author} {\bibinfo {author} {\bibfnamefont {M.}~\bibnamefont
  {Germano}},\ }\href {\doibase 10.1017/S0022112092001733} {\bibfield
  {journal} {\bibinfo  {journal} {J. Fluid. Mech.}\ }\textbf {\bibinfo {volume}
  {238}},\ \bibinfo {pages} {325} (\bibinfo {year} {1992})},\ \bibinfo {note}
  {publisher: Cambridge University Press}\BibitemShut {NoStop}%
\bibitem [{\citenamefont {Eyink}(2005)}]{eyink_locality_2005}%
  \BibitemOpen
  \bibfield  {author} {\bibinfo {author} {\bibfnamefont {G.~L.}\ \bibnamefont
  {Eyink}},\ }\href {\doibase 10.1016/j.physd.2005.05.018} {\bibfield
  {journal} {\bibinfo  {journal} {Physica D: Nonlinear Phenomena}\ }\textbf
  {\bibinfo {volume} {207}},\ \bibinfo {pages} {91} (\bibinfo {year}
  {2005})}\BibitemShut {NoStop}%
\bibitem [{\citenamefont {Aluie}(2017)}]{aluieMHD}%
  \BibitemOpen
  \bibfield  {author} {\bibinfo {author} {\bibfnamefont {H.}~\bibnamefont
  {Aluie}},\ }\href {\doibase 10.1088/1367-2630/aa5d2f} {\bibfield  {journal}
  {\bibinfo  {journal} {New Journal of Physics}\ }\textbf {\bibinfo {volume}
  {19}},\ \bibinfo {pages} {025008} (\bibinfo {year} {2017})}\BibitemShut
  {NoStop}%
\bibitem [{\citenamefont {Camporeale}\ \emph {et~al.}(2018)\citenamefont
  {Camporeale}, \citenamefont {Sorriso-Valvo}, \citenamefont {Califano},\ and\
  \citenamefont {Retin\`o}}]{CamporealePRL}%
  \BibitemOpen
  \bibfield  {author} {\bibinfo {author} {\bibfnamefont {E.}~\bibnamefont
  {Camporeale}}, \bibinfo {author} {\bibfnamefont {L.}~\bibnamefont
  {Sorriso-Valvo}}, \bibinfo {author} {\bibfnamefont {F.}~\bibnamefont
  {Califano}}, \ and\ \bibinfo {author} {\bibfnamefont {A.}~\bibnamefont
  {Retin\`o}},\ }\href {\doibase 10.1103/PhysRevLett.120.125101} {\bibfield
  {journal} {\bibinfo  {journal} {Phys. Rev. Lett.}\ }\textbf {\bibinfo
  {volume} {120}},\ \bibinfo {pages} {125101} (\bibinfo {year}
  {2018})}\BibitemShut {NoStop}%
\bibitem [{\citenamefont {Cerri}\ and\ \citenamefont
  {Camporeale}(2020)}]{Cerri_camporeale}%
  \BibitemOpen
  \bibfield  {author} {\bibinfo {author} {\bibfnamefont {S.~S.}\ \bibnamefont
  {Cerri}}\ and\ \bibinfo {author} {\bibfnamefont {E.}~\bibnamefont
  {Camporeale}},\ }\href {\doibase 10.1063/5.0012924} {\bibfield  {journal}
  {\bibinfo  {journal} {Phys. Plasmas}\ }\textbf {\bibinfo {volume} {27}},\
  \bibinfo {pages} {082102} (\bibinfo {year} {2020})},\ \Eprint
  {http://arxiv.org/abs/https://doi.org/10.1063/5.0012924}
  {https://doi.org/10.1063/5.0012924} \BibitemShut {NoStop}%
\bibitem [{\citenamefont {Yang}\ \emph {et~al.}(2017)\citenamefont {Yang},
  \citenamefont {Matthaeus}, \citenamefont {Parashar}, \citenamefont
  {Haggerty}, \citenamefont {Roytershteyn}, \citenamefont {Daughton},
  \citenamefont {Wan}, \citenamefont {Shi},\ and\ \citenamefont
  {Chen}}]{Yang17}%
  \BibitemOpen
  \bibfield  {author} {\bibinfo {author} {\bibfnamefont {Y.}~\bibnamefont
  {Yang}}, \bibinfo {author} {\bibfnamefont {W.~H.}\ \bibnamefont {Matthaeus}},
  \bibinfo {author} {\bibfnamefont {T.~N.}\ \bibnamefont {Parashar}}, \bibinfo
  {author} {\bibfnamefont {C.~C.}\ \bibnamefont {Haggerty}}, \bibinfo {author}
  {\bibfnamefont {V.}~\bibnamefont {Roytershteyn}}, \bibinfo {author}
  {\bibfnamefont {W.}~\bibnamefont {Daughton}}, \bibinfo {author}
  {\bibfnamefont {M.}~\bibnamefont {Wan}}, \bibinfo {author} {\bibfnamefont
  {Y.}~\bibnamefont {Shi}}, \ and\ \bibinfo {author} {\bibfnamefont
  {S.}~\bibnamefont {Chen}},\ }\href {\doibase 10.1063/1.4990421} {\bibfield
  {journal} {\bibinfo  {journal} {Phys. Plasmas}\ }\textbf {\bibinfo {volume}
  {24}},\ \bibinfo {pages} {072306} (\bibinfo {year} {2017})},\ \Eprint
  {http://arxiv.org/abs/https://doi.org/10.1063/1.4990421}
  {https://doi.org/10.1063/1.4990421} \BibitemShut {NoStop}%
\bibitem [{\citenamefont {Manzini}\ \emph {et~al.}(2022)\citenamefont
  {Manzini}, \citenamefont {Sahraoui}, \citenamefont {Califano},\ and\
  \citenamefont {~}}]{manzini2022}%
  \BibitemOpen
  \bibfield  {author} {\bibinfo {author} {\bibfnamefont {D.}~\bibnamefont
  {Manzini}}, \bibinfo {author} {\bibfnamefont {F.}~\bibnamefont {Sahraoui}},
  \bibinfo {author} {\bibfnamefont {F.}~\bibnamefont {Califano}}, \ and\
  \bibinfo {author} {\bibfnamefont {R.}~\bibnamefont {~}},\ }\href {\doibase
  10.1103/PhysRevE.106.035202} {\bibfield  {journal} {\bibinfo  {journal}
  {Phys. Rev. E}\ }\textbf {\bibinfo {volume} {106}},\ \bibinfo {pages}
  {035202} (\bibinfo {year} {2022})}\BibitemShut {NoStop}%
\bibitem [{\citenamefont {Manzini}\ \emph {et~al.}(2023)\citenamefont
  {Manzini}, \citenamefont {Sahraoui},\ and\ \citenamefont
  {Califano}}]{manzini23}%
  \BibitemOpen
  \bibfield  {author} {\bibinfo {author} {\bibfnamefont {D.}~\bibnamefont
  {Manzini}}, \bibinfo {author} {\bibfnamefont {F.}~\bibnamefont {Sahraoui}}, \
  and\ \bibinfo {author} {\bibfnamefont {F.}~\bibnamefont {Califano}},\ }\href
  {\doibase 10.1103/PhysRevLett.130.205201} {\bibfield  {journal} {\bibinfo
  {journal} {Phys. Rev. Lett.}\ }\textbf {\bibinfo {volume} {130}},\ \bibinfo
  {pages} {205201} (\bibinfo {year} {2023})}\BibitemShut {NoStop}%
\bibitem [{\citenamefont {Adhikari}\ \emph {et~al.}(2023)\citenamefont
  {Adhikari}, \citenamefont {Yang}, \citenamefont {Matthaeus}, \citenamefont
  {Cassak}, \citenamefont {Parashar},\ and\ \citenamefont
  {Shay}}]{adhikari2023scale}%
  \BibitemOpen
  \bibfield  {author} {\bibinfo {author} {\bibfnamefont {S.}~\bibnamefont
  {Adhikari}}, \bibinfo {author} {\bibfnamefont {Y.}~\bibnamefont {Yang}},
  \bibinfo {author} {\bibfnamefont {W.~H.}\ \bibnamefont {Matthaeus}}, \bibinfo
  {author} {\bibfnamefont {P.~A.}\ \bibnamefont {Cassak}}, \bibinfo {author}
  {\bibfnamefont {T.~N.}\ \bibnamefont {Parashar}}, \ and\ \bibinfo {author}
  {\bibfnamefont {M.~A.}\ \bibnamefont {Shay}},\ }\href@noop {} {\enquote
  {\bibinfo {title} {Scale filtering analysis of kinetic reconnection and its
  associated turbulence},}\ } (\bibinfo {year} {2023}),\ \Eprint
  {http://arxiv.org/abs/2310.16973} {arXiv:2310.16973 [physics.plasm-ph]}
  \BibitemShut {NoStop}%
\bibitem [{\citenamefont {Belmont}\ \emph {et~al.}(2013)\citenamefont
  {Belmont}, \citenamefont {Grappin}, \citenamefont {Mottez}, \citenamefont
  {Pantellini},\ and\ \citenamefont {Pelletier}}]{Belmont_book}%
  \BibitemOpen
  \bibfield  {author} {\bibinfo {author} {\bibfnamefont {G.}~\bibnamefont
  {Belmont}}, \bibinfo {author} {\bibfnamefont {R.}~\bibnamefont {Grappin}},
  \bibinfo {author} {\bibfnamefont {F.}~\bibnamefont {Mottez}}, \bibinfo
  {author} {\bibfnamefont {F.}~\bibnamefont {Pantellini}}, \ and\ \bibinfo
  {author} {\bibfnamefont {G.}~\bibnamefont {Pelletier}},\ }\href {\doibase
  10.1002/9783527656226} {\emph {\bibinfo {title} {Collisionless Plasmas in
  Astrophysics}}}\ (\bibinfo {year} {2013})\ p.\ \bibinfo {pages}
  {119}\BibitemShut {NoStop}%
\bibitem [{\citenamefont {Fitzpatrick}(2022)}]{Fiitzpatrick_book}%
  \BibitemOpen
  \bibfield  {author} {\bibinfo {author} {\bibfnamefont {R.}~\bibnamefont
  {Fitzpatrick}},\ }\href {\doibase 10.1201/9781003268253} {\emph {\bibinfo
  {title} {Plasma Physics. An Introduction}}}\ (\bibinfo {year}
  {2022})\BibitemShut {NoStop}%
\bibitem [{\citenamefont {Aluie}(2011)}]{aluie_compressible_2011}%
  \BibitemOpen
  \bibfield  {author} {\bibinfo {author} {\bibfnamefont {H.}~\bibnamefont
  {Aluie}},\ }\href {\doibase 10.1103/PhysRevLett.106.174502} {\bibfield
  {journal} {\bibinfo  {journal} {Phys. Rev. Lett.}\ }\textbf {\bibinfo
  {volume} {106}},\ \bibinfo {pages} {174502} (\bibinfo {year}
  {2011})}\BibitemShut {NoStop}%
\bibitem [{\citenamefont {Aluie}(2013)}]{Aluie_compressible}%
  \BibitemOpen
  \bibfield  {author} {\bibinfo {author} {\bibfnamefont {H.}~\bibnamefont
  {Aluie}},\ }\href {\doibase https://doi.org/10.1016/j.physd.2012.12.009}
  {\bibfield  {journal} {\bibinfo  {journal} {Physica D: Nonlinear Phenomena}\
  }\textbf {\bibinfo {volume} {247}},\ \bibinfo {pages} {54} (\bibinfo {year}
  {2013})}\BibitemShut {NoStop}%
\bibitem [{\citenamefont {Howes}\ \emph {et~al.}(2017)\citenamefont {Howes},
  \citenamefont {Klein},\ and\ \citenamefont {Li}}]{howes_klein_li_2017}%
  \BibitemOpen
  \bibfield  {author} {\bibinfo {author} {\bibfnamefont {G.~G.}\ \bibnamefont
  {Howes}}, \bibinfo {author} {\bibfnamefont {K.~G.}\ \bibnamefont {Klein}}, \
  and\ \bibinfo {author} {\bibfnamefont {T.~C.}\ \bibnamefont {Li}},\ }\href
  {\doibase 10.1017/S0022377816001197} {\bibfield  {journal} {\bibinfo
  {journal} {J. Plasma Phys.}\ }\textbf {\bibinfo {volume} {83}},\ \bibinfo
  {pages} {705830102} (\bibinfo {year} {2017})}\BibitemShut {NoStop}%
\bibitem [{\citenamefont {Afshari}\ \emph {et~al.}(2021)\citenamefont
  {Afshari}, \citenamefont {Howes}, \citenamefont {Kletzing}, \citenamefont
  {Hartley},\ and\ \citenamefont {Boardsen}}]{afshari_ELD}%
  \BibitemOpen
  \bibfield  {author} {\bibinfo {author} {\bibfnamefont {A.~S.}\ \bibnamefont
  {Afshari}}, \bibinfo {author} {\bibfnamefont {G.~G.}\ \bibnamefont {Howes}},
  \bibinfo {author} {\bibfnamefont {C.~A.}\ \bibnamefont {Kletzing}}, \bibinfo
  {author} {\bibfnamefont {D.~P.}\ \bibnamefont {Hartley}}, \ and\ \bibinfo
  {author} {\bibfnamefont {S.~A.}\ \bibnamefont {Boardsen}},\ }\href {\doibase
  https://doi.org/10.1029/2021JA029578} {\bibfield  {journal} {\bibinfo
  {journal} {Journal of Geophysical Research: Space Physics}\ }\textbf
  {\bibinfo {volume} {126}},\ \bibinfo {pages} {e2021JA029578} (\bibinfo {year}
  {2021})},\ \bibinfo {note} {e2021JA029578 2021JA029578}\BibitemShut {NoStop}%
\bibitem [{\citenamefont {Ferrand}\ \emph {et~al.}(2021)\citenamefont
  {Ferrand}, \citenamefont {Sahraoui}, \citenamefont {Laveder}, \citenamefont
  {Passot}, \citenamefont {Sulem},\ and\ \citenamefont
  {Galtier}}]{Ferrand_2021LF}%
  \BibitemOpen
  \bibfield  {author} {\bibinfo {author} {\bibfnamefont {R.}~\bibnamefont
  {Ferrand}}, \bibinfo {author} {\bibfnamefont {F.}~\bibnamefont {Sahraoui}},
  \bibinfo {author} {\bibfnamefont {D.}~\bibnamefont {Laveder}}, \bibinfo
  {author} {\bibfnamefont {T.}~\bibnamefont {Passot}}, \bibinfo {author}
  {\bibfnamefont {P.~L.}\ \bibnamefont {Sulem}}, \ and\ \bibinfo {author}
  {\bibfnamefont {S.}~\bibnamefont {Galtier}},\ }\href {\doibase
  10.3847/1538-4357/ac2bfb} {\bibfield  {journal} {\bibinfo  {journal}
  {Astrophys. J.}\ }\textbf {\bibinfo {volume} {923}},\ \bibinfo {pages} {122}
  (\bibinfo {year} {2021})}\BibitemShut {NoStop}%
\bibitem [{\citenamefont {Burch}\ \emph {et~al.}(2016)\citenamefont {Burch},
  \citenamefont {Moore}, \citenamefont {Torbert},\ and\ \citenamefont
  {Giles}}]{Burch2016}%
  \BibitemOpen
  \bibfield  {author} {\bibinfo {author} {\bibfnamefont {J.~L.}\ \bibnamefont
  {Burch}}, \bibinfo {author} {\bibfnamefont {T.~E.}\ \bibnamefont {Moore}},
  \bibinfo {author} {\bibfnamefont {R.~B.}\ \bibnamefont {Torbert}}, \ and\
  \bibinfo {author} {\bibfnamefont {B.~L.}\ \bibnamefont {Giles}},\ }\href
  {\doibase 10.1007/s11214-015-0164-9} {\bibfield  {journal} {\bibinfo
  {journal} {Space Science Reviews}\ }\textbf {\bibinfo {volume} {199}},\
  \bibinfo {pages} {5} (\bibinfo {year} {2016})}\BibitemShut {NoStop}%
\bibitem [{\citenamefont {Chanteur}(1998)}]{chanteur_spatial_1998}%
  \BibitemOpen
  \bibfield  {author} {\bibinfo {author} {\bibfnamefont {G.}~\bibnamefont
  {Chanteur}},\ }\href {https://ui.adsabs.harvard.edu/abs/1998ISSIR...1..349C}
  {\bibfield  {journal} {\bibinfo  {journal} {{ISSI} Scientific Reports
  Series}\ }\textbf {\bibinfo {volume} {1}},\ \bibinfo {pages} {349} (\bibinfo
  {year} {1998})}\BibitemShut {NoStop}%
\bibitem [{\citenamefont {Lindqvist}\ \emph {et~al.}(2016)\citenamefont
  {Lindqvist}, \citenamefont {Olsson}, \citenamefont {Torbert}, \citenamefont
  {King}, \citenamefont {Granoff} \emph {et~al.}}]{lindqvist_spin-plane_2016}%
  \BibitemOpen
  \bibfield  {author} {\bibinfo {author} {\bibfnamefont {P.-A.}\ \bibnamefont
  {Lindqvist}}, \bibinfo {author} {\bibfnamefont {G.}~\bibnamefont {Olsson}},
  \bibinfo {author} {\bibfnamefont {R.~B.}\ \bibnamefont {Torbert}}, \bibinfo
  {author} {\bibfnamefont {B.}~\bibnamefont {King}}, \bibinfo {author}
  {\bibfnamefont {M.}~\bibnamefont {Granoff}},  \emph {et~al.},\ }\href
  {\doibase 10.1007/s11214-014-0116-9} {\bibfield  {journal} {\bibinfo
  {journal} {Space Sci. Rev.}\ }\textbf {\bibinfo {volume} {199}},\ \bibinfo
  {pages} {137} (\bibinfo {year} {2016})}\BibitemShut {NoStop}%
\bibitem [{\citenamefont {Ergun}\ \emph {et~al.}(2016)\citenamefont {Ergun},
  \citenamefont {Tucker}, \citenamefont {Westfall}, \citenamefont {Goodrich},
  \citenamefont {Malaspina} \emph {et~al.}}]{ergun_axial_2016}%
  \BibitemOpen
  \bibfield  {author} {\bibinfo {author} {\bibfnamefont {R.~E.}\ \bibnamefont
  {Ergun}}, \bibinfo {author} {\bibfnamefont {S.}~\bibnamefont {Tucker}},
  \bibinfo {author} {\bibfnamefont {J.}~\bibnamefont {Westfall}}, \bibinfo
  {author} {\bibfnamefont {K.~A.}\ \bibnamefont {Goodrich}}, \bibinfo {author}
  {\bibfnamefont {D.~M.}\ \bibnamefont {Malaspina}},  \emph {et~al.},\ }\href
  {\doibase 10.1007/s11214-014-0115-x} {\bibfield  {journal} {\bibinfo
  {journal} {Space Sci. Rev.}\ }\textbf {\bibinfo {volume} {199}},\ \bibinfo
  {pages} {167} (\bibinfo {year} {2016})}\BibitemShut {NoStop}%
\bibitem [{\citenamefont {Pollock}\ \emph {et~al.}(2016)\citenamefont
  {Pollock}, \citenamefont {Moore}, \citenamefont {Jacques}, \citenamefont
  {Burch}, \citenamefont {Gliese} \emph {et~al.}}]{pollock_fast_2016}%
  \BibitemOpen
  \bibfield  {author} {\bibinfo {author} {\bibfnamefont {C.}~\bibnamefont
  {Pollock}}, \bibinfo {author} {\bibfnamefont {T.}~\bibnamefont {Moore}},
  \bibinfo {author} {\bibfnamefont {A.}~\bibnamefont {Jacques}}, \bibinfo
  {author} {\bibfnamefont {J.}~\bibnamefont {Burch}}, \bibinfo {author}
  {\bibfnamefont {U.}~\bibnamefont {Gliese}},  \emph {et~al.},\ }\href
  {\doibase 10.1007/s11214-016-0245-4} {\bibfield  {journal} {\bibinfo
  {journal} {Space Sci. Rev.}\ }\textbf {\bibinfo {volume} {199}},\ \bibinfo
  {pages} {331} (\bibinfo {year} {2016})}\BibitemShut {NoStop}%
\bibitem [{\citenamefont {Welch}(1967)}]{Welch}%
  \BibitemOpen
  \bibfield  {author} {\bibinfo {author} {\bibfnamefont {P.}~\bibnamefont
  {Welch}},\ }\href {\doibase 10.1109/TAU.1967.1161901} {\bibfield  {journal}
  {\bibinfo  {journal} {IEEE Transactions on Audio and Electroacoustics}\
  }\textbf {\bibinfo {volume} {15}},\ \bibinfo {pages} {70} (\bibinfo {year}
  {1967})}\BibitemShut {NoStop}%
\bibitem [{\citenamefont {Hellinger}\ \emph {et~al.}(2022)\citenamefont
  {Hellinger}, \citenamefont {Montagud-Camps}, \citenamefont {Franci},
  \citenamefont {Matteini}, \citenamefont {Papini}, \citenamefont {Verdini},\
  and\ \citenamefont {Landi}}]{Hellinger_2022}%
  \BibitemOpen
  \bibfield  {author} {\bibinfo {author} {\bibfnamefont {P.}~\bibnamefont
  {Hellinger}}, \bibinfo {author} {\bibfnamefont {V.}~\bibnamefont
  {Montagud-Camps}}, \bibinfo {author} {\bibfnamefont {L.}~\bibnamefont
  {Franci}}, \bibinfo {author} {\bibfnamefont {L.}~\bibnamefont {Matteini}},
  \bibinfo {author} {\bibfnamefont {E.}~\bibnamefont {Papini}}, \bibinfo
  {author} {\bibfnamefont {A.}~\bibnamefont {Verdini}}, \ and\ \bibinfo
  {author} {\bibfnamefont {S.}~\bibnamefont {Landi}},\ }\href {\doibase
  10.3847/1538-4357/ac5fad} {\bibfield  {journal} {\bibinfo  {journal}
  {Astrophys. J.}\ }\textbf {\bibinfo {volume} {930}},\ \bibinfo {pages} {48}
  (\bibinfo {year} {2022})}\BibitemShut {NoStop}%
\bibitem [{\citenamefont {Yang}\ \emph {et~al.}(2022)\citenamefont {Yang},
  \citenamefont {Matthaeus}, \citenamefont {Roy}, \citenamefont {Roytershteyn},
  \citenamefont {Parashar}, \citenamefont {Bandyopadhyay},\ and\ \citenamefont
  {Wan}}]{Yang_2022}%
  \BibitemOpen
  \bibfield  {author} {\bibinfo {author} {\bibfnamefont {Y.}~\bibnamefont
  {Yang}}, \bibinfo {author} {\bibfnamefont {W.~H.}\ \bibnamefont {Matthaeus}},
  \bibinfo {author} {\bibfnamefont {S.}~\bibnamefont {Roy}}, \bibinfo {author}
  {\bibfnamefont {V.}~\bibnamefont {Roytershteyn}}, \bibinfo {author}
  {\bibfnamefont {T.~N.}\ \bibnamefont {Parashar}}, \bibinfo {author}
  {\bibfnamefont {R.}~\bibnamefont {Bandyopadhyay}}, \ and\ \bibinfo {author}
  {\bibfnamefont {M.}~\bibnamefont {Wan}},\ }\href {\doibase
  10.3847/1538-4357/ac5d3e} {\bibfield  {journal} {\bibinfo  {journal} {The
  Astrophysical Journal}\ }\textbf {\bibinfo {volume} {929}},\ \bibinfo {pages}
  {142} (\bibinfo {year} {2022})}\BibitemShut {NoStop}%
\bibitem [{\citenamefont {Roy}\ \emph {et~al.}(2022)\citenamefont {Roy},
  \citenamefont {Bandyopadhyay}, \citenamefont {Yang}, \citenamefont
  {Parashar}, \citenamefont {Matthaeus}, \citenamefont {Adhikari},
  \citenamefont {Roytershteyn}, \citenamefont {Chasapis}, \citenamefont {Li},
  \citenamefont {Gershman}, \citenamefont {Giles},\ and\ \citenamefont
  {Burch}}]{Roy_2022}%
  \BibitemOpen
  \bibfield  {author} {\bibinfo {author} {\bibfnamefont {S.}~\bibnamefont
  {Roy}}, \bibinfo {author} {\bibfnamefont {R.}~\bibnamefont {Bandyopadhyay}},
  \bibinfo {author} {\bibfnamefont {Y.}~\bibnamefont {Yang}}, \bibinfo {author}
  {\bibfnamefont {T.~N.}\ \bibnamefont {Parashar}}, \bibinfo {author}
  {\bibfnamefont {W.~H.}\ \bibnamefont {Matthaeus}}, \bibinfo {author}
  {\bibfnamefont {S.}~\bibnamefont {Adhikari}}, \bibinfo {author}
  {\bibfnamefont {V.}~\bibnamefont {Roytershteyn}}, \bibinfo {author}
  {\bibfnamefont {A.}~\bibnamefont {Chasapis}}, \bibinfo {author}
  {\bibfnamefont {H.}~\bibnamefont {Li}}, \bibinfo {author} {\bibfnamefont
  {D.~J.}\ \bibnamefont {Gershman}}, \bibinfo {author} {\bibfnamefont {B.~L.}\
  \bibnamefont {Giles}}, \ and\ \bibinfo {author} {\bibfnamefont {J.~L.}\
  \bibnamefont {Burch}},\ }\href {\doibase 10.3847/1538-4357/aca479} {\bibfield
   {journal} {\bibinfo  {journal} {Astrophys. J.}\ }\textbf {\bibinfo {volume}
  {941}},\ \bibinfo {pages} {137} (\bibinfo {year} {2022})}\BibitemShut
  {NoStop}%
\bibitem [{\citenamefont {Sahraoui}\ \emph {et~al.}(2009)\citenamefont
  {Sahraoui}, \citenamefont {Goldstein}, \citenamefont {Robert},\ and\
  \citenamefont {Khotyaintsev}}]{sahraoui2009}%
  \BibitemOpen
  \bibfield  {author} {\bibinfo {author} {\bibfnamefont {F.}~\bibnamefont
  {Sahraoui}}, \bibinfo {author} {\bibfnamefont {M.~L.}\ \bibnamefont
  {Goldstein}}, \bibinfo {author} {\bibfnamefont {P.}~\bibnamefont {Robert}}, \
  and\ \bibinfo {author} {\bibfnamefont {Y.~V.}\ \bibnamefont {Khotyaintsev}},\
  }\href {\doibase 10.1103/PhysRevLett.102.231102} {\bibfield  {journal}
  {\bibinfo  {journal} {Phys. Rev. Lett.}\ }\textbf {\bibinfo {volume} {102}},\
  \bibinfo {pages} {231102} (\bibinfo {year} {2009})}\BibitemShut {NoStop}%
\bibitem [{\citenamefont {Howes}\ \emph {et~al.}(2011)\citenamefont {Howes},
  \citenamefont {TenBarge}, \citenamefont {Dorland}, \citenamefont {Quataert},
  \citenamefont {Schekochihin}, \citenamefont {Numata},\ and\ \citenamefont
  {Tatsuno}}]{howes2011}%
  \BibitemOpen
  \bibfield  {author} {\bibinfo {author} {\bibfnamefont {G.~G.}\ \bibnamefont
  {Howes}}, \bibinfo {author} {\bibfnamefont {J.~M.}\ \bibnamefont {TenBarge}},
  \bibinfo {author} {\bibfnamefont {W.}~\bibnamefont {Dorland}}, \bibinfo
  {author} {\bibfnamefont {E.}~\bibnamefont {Quataert}}, \bibinfo {author}
  {\bibfnamefont {A.~A.}\ \bibnamefont {Schekochihin}}, \bibinfo {author}
  {\bibfnamefont {R.}~\bibnamefont {Numata}}, \ and\ \bibinfo {author}
  {\bibfnamefont {T.}~\bibnamefont {Tatsuno}},\ }\href {\doibase
  10.1103/PhysRevLett.107.035004} {\bibfield  {journal} {\bibinfo  {journal}
  {Phys. Rev. Lett.}\ }\textbf {\bibinfo {volume} {107}},\ \bibinfo {pages}
  {035004} (\bibinfo {year} {2011})}\BibitemShut {NoStop}%
\bibitem [{\citenamefont {Sahraoui}\ \emph {et~al.}(2010)\citenamefont
  {Sahraoui}, \citenamefont {Goldstein}, \citenamefont {Belmont}, \citenamefont
  {Canu},\ and\ \citenamefont {Rezeau}}]{sahraoui2010}%
  \BibitemOpen
  \bibfield  {author} {\bibinfo {author} {\bibfnamefont {F.}~\bibnamefont
  {Sahraoui}}, \bibinfo {author} {\bibfnamefont {M.~L.}\ \bibnamefont
  {Goldstein}}, \bibinfo {author} {\bibfnamefont {G.}~\bibnamefont {Belmont}},
  \bibinfo {author} {\bibfnamefont {P.}~\bibnamefont {Canu}}, \ and\ \bibinfo
  {author} {\bibfnamefont {L.}~\bibnamefont {Rezeau}},\ }\href {\doibase
  10.1103/PhysRevLett.105.131101} {\bibfield  {journal} {\bibinfo  {journal}
  {Phys. Rev. Lett.}\ }\textbf {\bibinfo {volume} {105}},\ \bibinfo {pages}
  {131101} (\bibinfo {year} {2010})}\BibitemShut {NoStop}%
\bibitem [{\citenamefont {Schekochihin}\ \emph {et~al.}(2016)\citenamefont
  {Schekochihin}, \citenamefont {Parker}, \citenamefont {Highcock},
  \citenamefont {Dellar}, \citenamefont {Dorland},\ and\ \citenamefont
  {Hammett}}]{schekochihinHermite}%
  \BibitemOpen
  \bibfield  {author} {\bibinfo {author} {\bibfnamefont {A.~A.}\ \bibnamefont
  {Schekochihin}}, \bibinfo {author} {\bibfnamefont {J.~T.}\ \bibnamefont
  {Parker}}, \bibinfo {author} {\bibfnamefont {E.~G.}\ \bibnamefont
  {Highcock}}, \bibinfo {author} {\bibfnamefont {P.~J.}\ \bibnamefont
  {Dellar}}, \bibinfo {author} {\bibfnamefont {W.}~\bibnamefont {Dorland}}, \
  and\ \bibinfo {author} {\bibfnamefont {G.~W.}\ \bibnamefont {Hammett}},\
  }\href {\doibase 10.1017/S0022377816000374} {\bibfield  {journal} {\bibinfo
  {journal} {Journal of Plasma Physics}\ }\textbf {\bibinfo {volume} {82}},\
  \bibinfo {pages} {905820212} (\bibinfo {year} {2016})}\BibitemShut {NoStop}%
\bibitem [{\citenamefont {Servidio}\ \emph {et~al.}(2017)\citenamefont
  {Servidio}, \citenamefont {Chasapis}, \citenamefont {Matthaeus},
  \citenamefont {Perrone}, \citenamefont {Valentini}, \citenamefont {Parashar},
  \citenamefont {Veltri}, \citenamefont {Gershman}, \citenamefont {Russell},
  \citenamefont {Giles}, \citenamefont {Fuselier}, \citenamefont {Phan},\ and\
  \citenamefont {Burch}}]{servidioHermite}%
  \BibitemOpen
  \bibfield  {author} {\bibinfo {author} {\bibfnamefont {S.}~\bibnamefont
  {Servidio}}, \bibinfo {author} {\bibfnamefont {A.}~\bibnamefont {Chasapis}},
  \bibinfo {author} {\bibfnamefont {W.~H.}\ \bibnamefont {Matthaeus}}, \bibinfo
  {author} {\bibfnamefont {D.}~\bibnamefont {Perrone}}, \bibinfo {author}
  {\bibfnamefont {F.}~\bibnamefont {Valentini}}, \bibinfo {author}
  {\bibfnamefont {T.~N.}\ \bibnamefont {Parashar}}, \bibinfo {author}
  {\bibfnamefont {P.}~\bibnamefont {Veltri}}, \bibinfo {author} {\bibfnamefont
  {D.}~\bibnamefont {Gershman}}, \bibinfo {author} {\bibfnamefont {C.~T.}\
  \bibnamefont {Russell}}, \bibinfo {author} {\bibfnamefont {B.}~\bibnamefont
  {Giles}}, \bibinfo {author} {\bibfnamefont {S.~A.}\ \bibnamefont {Fuselier}},
  \bibinfo {author} {\bibfnamefont {T.~D.}\ \bibnamefont {Phan}}, \ and\
  \bibinfo {author} {\bibfnamefont {J.}~\bibnamefont {Burch}},\ }\href
  {\doibase 10.1103/PhysRevLett.119.205101} {\bibfield  {journal} {\bibinfo
  {journal} {Phys. Rev. Lett.}\ }\textbf {\bibinfo {volume} {119}},\ \bibinfo
  {pages} {205101} (\bibinfo {year} {2017})}\BibitemShut {NoStop}%
\bibitem [{\citenamefont {Cassak}\ \emph {et~al.}(2023)\citenamefont {Cassak},
  \citenamefont {Barbhuiya}, \citenamefont {Liang},\ and\ \citenamefont
  {Argall}}]{cassak_2023}%
  \BibitemOpen
  \bibfield  {author} {\bibinfo {author} {\bibfnamefont {P.~A.}\ \bibnamefont
  {Cassak}}, \bibinfo {author} {\bibfnamefont {M.~H.}\ \bibnamefont
  {Barbhuiya}}, \bibinfo {author} {\bibfnamefont {H.}~\bibnamefont {Liang}}, \
  and\ \bibinfo {author} {\bibfnamefont {M.~R.}\ \bibnamefont {Argall}},\
  }\href {\doibase 10.1103/PhysRevLett.130.085201} {\bibfield  {journal}
  {\bibinfo  {journal} {Phys. Rev. Lett.}\ }\textbf {\bibinfo {volume} {130}},\
  \bibinfo {pages} {085201} (\bibinfo {year} {2023})}\BibitemShut {NoStop}%
\end{thebibliography}

\end{document}